\documentclass[sigconf]{acmart}

\usepackage{graphicx}
\usepackage{multirow}
\usepackage{array}

\AtBeginDocument{%
  \providecommand\BibTeX{{%
    \normalfont B\kern-0.5em{\scshape i\kern-0.25em b}\kern-0.8em\TeX}}}

\copyrightyear{2019}
\acmYear{2019}
\setcopyright{rightsretained}

\acmConference{ICSE'42}{May 23 - 29, 2020}{Seoul, South Korea}
\acmDOI{***}
\acmISBN{***}
\acmBooktitle{ICSE'42, May 23 - 29, 2020, Seoul, South Korea}

\begin{document}

\title{Software Engineering Practice in the Development of \\ Deep Learning Applications}

\author{Xufan Zhang}
\affiliation{
  \institution{The State Key Laboratory of Novel Software Technology}
  \city{Nanjing} 
  \state{Jiangsu} 
  \country{China}
}
\email{xufan.zhang@outlook.com}

\author{Yilin Yang}
\affiliation{
  \institution{The State Key Laboratory of Novel Software Technology}
  \city{Nanjing} 
  \state{Jiangsu} 
  \country{China}
}
\email{DG1732004@smail.nju.edu.cn}

\author{Yang Feng}
\authornote{Yang Feng and Zhenyu Chen are the corresponding authors.}
\authornotemark[0]
\affiliation{
  \institution{The State Key Laboratory of Novel Software Technology}
  \city{Nanjing} 
  \state{Jiangsu} 
  \country{China}
}
\email{charles.fengy@gmail.com}

\author{Zhenyu Chen}
\authornotemark[1]
\affiliation{
  \institution{The State Key Laboratory of Novel Software Technology}
  \city{Nanjing} 
  \state{Jiangsu} 
  \country{China}
}
\email{zychen@nju.edu.cn}


\begin{abstract}

Deep-Learning(DL) applications have been widely employed to assist in various tasks.
They are constructed based on a data-driven programming paradigm that is different from conventional software applications. 
Given the increasing popularity and importance of DL applications, software engineering practitioners have some techniques specifically for them.
However, little research is conducted to identify the challenges and lacks in practice.
To fill this gap, in this paper, we surveyed 195 practitioners to understand their insight and experience in the software engineering practice of DL applications.
Specifically, we asked the respondents to identify lacks and challenges in the practice of the development life cycle of DL applications.
The results present 13 findings that provide us with a better understanding of software engineering practice of DL applications. 
Further, we distil these findings into 7 actionable recommendations for software engineering researchers and practitioners to improve the development of DL applications.

\end{abstract}


\maketitle

\section{Introduction}

The tremendous advancement of deep learning(DL) techniques has driven the emergence of DL applications that offer commercial benefits to humans, and they are gradually deployed in more life-critical fields, such as medical diagnosis~\cite{gulshan2016development,wong2016artificial}, in autonomous vehicles~\cite{bojarski_end_2016,huval2015empirical}, and air traffic control~\cite{katz2017reluplex}. 
Under this circumstance, the development of DL applications has become vital for their success. 
However, the development of DL applications adopts a programming paradigm and practice that is entirely different from conventional software applications~\cite{zhang2019machine,amershi2019software,zhangempirical}. 
Therefore, methods, metrics, and techniques that are designed for the development of conventional software applications may become less effective to be applied to the development of DL applications. 

To facilitate the development of DL applications, software engineering researchers have proposed several metrics~\cite{ma2019deepct,ma2018deepgauge,ma2018deepmutation,sun2018testing} and techniques~\cite{tian2018deeptest,pei2017deepxplore,xie2019deephunter,guo2018dlfuzz,du2019deepstellar}, however, it is unclear what kinds of lacks and challenges practitioners face in developing DL applications. 
Understanding the software engineering practices of developing DL applications is the first, yet critical, step towards building useful and effective techniques.  Several prior research has been conducted to investigate the challenges in developing DL applications~\cite{zhangempirical,wan2019does} and also to characterize the bugs of DL frameworks~\cite{zhangempirical2018,islam2019comprehensive}. Zhang et al.~\cite{zhangempirical} conduct an empirical study on the deep learning questions on Stack Overflow and present a classification model to quantify the distribution of different kinds of deep learning questions. Their study reveals the common root causes of bugs and the most frequently asked questions in building applications based on the DL frameworks. Wan et al.~\cite{wan2019does} performed a mixture of qualitative and quantitative studies to investigate the differences in software practices and practitioners' work due to the impact of machine learning. Zhang et al.~\cite{zhangempirical2018} have conducted the first study to investigate the characteristics of the Tensorflow. They manually inspect and reproduce 75 Tensorflow bug reports from Github and 76 Tensorflow bug reports from StackOverflow. The study investigates the symptoms and common bug types of the Tensorflow framework. Also, they identify 5 challenges for researchers. 
Islam et al.\cite{islam2019comprehensive} inspect high-quality posts related to deep learning libraries on Stack Overflow and Github. 
They summarize the types of bugs, root causes of bugs, effects of bugs, bug-prone stage of deep learning pipeline as well as whether there are some common anti-patterns found in this buggy software. 
Even though this research have laid the first step towards understanding the new development paradigm, little research is conducted to understand the insights, experience, and expectations from the practitioners' perspective. 

In this study, we complement existing empirical studies by conducting a comprehensive survey with 195 industrial practitioners to understand the characteristics of each phase of the DL application development. 
We first conduct a literature review to understand the current software engineering studies on the DL application development. 
And then, we conduct interviews with eight developers from Baidu, Alibaba, and Huawei. 
Based on the literature review and interviews, we summarize the scope and focuses on a questionnaire. 
Finally, we distributed this questionnaire to practitioners from various companies with different backgrounds to provide feedback and opinions. 
We received a total of 221 responses, 195 of them are valid. 
Specifically, we also invited respondents to provide their rationale for the two hottest topics, i.e., \textit{testing} and debugging. 
These findings and feedback provide us with a comprehensive understanding of the vision and challenges of the DL application development. 

The main contributions of our work are as follows: 
\begin{itemize} 
\item We conduct a comprehensive survey with 195 practitioners to investigate the software engineering practice of developing DL applications. 

\item We summarize the results of this survey into 13 findings, which present the insight, experience, and expectations of practitioners. Also, these findings reveal the impacts and challenges in all phases of the DL application development life cycle. 

\item We distill these findings into 7 actionable recommendations for software engineering researchers. These recommendations can give researchers insight into designing various techniques for improving the development of DL applications. 
\end{itemize}

\section{Methodology}


Our study consists of three parts: 
1. a literature review is conducted to identify current software engineering research topics on the development of deep learning applications; 2. eight interviews with practitioners from Baidu, Alibaba, and Huawei are conducted to obtain insights into practitioner views to help us formulate a set of hypotheses;
3. a questionnaire summarized from the previous two steps is distributed to 822 practitioners with different backgrounds.

\subsection{Literature Review}



To understand the state-of-the-arts on the development of DL applications, we conducted a literature review. 
We first pick up 10 software engineering conferences, including ICSE, FSE, ASE, ICSME, MSR, SANER, ESEM, ICPC, ISSTA, ICST, and 7 journals, including  TOSEM, TSE, EMSE, ASE, JSS, IST, to collect relevant papers. 
We search the keywords "machine learning", "deep learning" and "deep neural network" in the paper title and abstract. 
Note that goal of this study is to identify challenges or problems in developing DL applications but not applying DL techniques to address software engineering problems. 
To ensure the paper's topic fits our goal, the first three authors read the abstract of each search result in the process. 

Finally, we obtained 16 papers, and summarize them as follows. 

\textbf{Requirement Analysis:} 
There is only one paper to discuss the requirement analysis in developing deep learning applications. 
Belan et al.~\cite{belani2019requirements} analyzed the related work from software engineering and AI fields that deal with requirements suitable for specifying a machine learning-based software solution. This work contributes to agent-based software engineering, goal-oriented requirements engineering, and practices for product development in companies. 

\textbf{Testing \& Debugging:}
To improve the quality of intelligent applications, researchers have cast substantial efforts on proposing specific testing and debugging techniques~\cite{ma2018deepgauge, sun2018concolic, xie2019deephunter, ma2018deepmutation, tiandeeptest2018, zhang2018road, pei2017deepxplore, guo2018dlfuzz}. 
Pei et al.\cite{pei2017deepxplore} designed, implemented and evaluated DeepXplore, the first white-box framework for system testing of real DL systems. 
They introduce neuron coverage to systematically measure portions of the DL system that are run by the test input and leverage multiple DL systems with similar functionality to cross-reference gods to avoid manual inspection. 
Tian et al.\cite{tiandeeptest2018} designed and developed DeepTest, a system testing tool for automatically detecting the wrong behavior of DNN-driven vehicles that can cause fatal collisions. 
Ma et al.\cite{ma2018deepgauge} proposed DeepGauge, a set of multi-granularity testing standards for DL systems designed to portray test platforms in many ways. Their in-depth evaluation of test standards was demonstrated in two well-known data sets, five DL systems, and four state-of-the-art countermeasures against DL. 
Sun et al.\cite{sun2018concolic} first introduced an analytical test method for deep neural networks (DNN). 
Ma et al.~\cite{ma2018deepmutation} presented a mutation testing framework for DNNs aiming at evaluating the quality of datasets. 
Guo et al.\cite{guo2018dlfuzz} proposed DLFuzz, the first differential fuzzy testing framework to guide the DL system to expose incorrect behavior. DLFuzz continually changes the input to maximize neuronal coverage and prediction differences between raw input and mutated input, without the need to manually mark work or cross-reference oracles of other DL systems with the same functionality. 
Xie et al.~\cite{xie2019deephunter} presented an automated fuzz testing framework, namely DeepHunter, for hunting potential defects of general-purpose DNNs. DeepHunter is designed based on metamorphic testing. It generates new semantically preserved tests, and leverages multiple plugable coverage criteria as feedback to guide the test generation from different perspectives.

\textbf{Empirical studies on characterizing the development of DL applications: }
Zhang\cite{zhangempirical2018} et al. researched the program bug of Tensorflow. They filtered the collected information from GitHub and StackOverflow, manually obtained more than 100 fine samples. Through analysis, they summarized three symptoms of symptoms (Symptoms) and six root causes. 
Wan et al.~\cite{wan2019does} studied the features and impacts of machine learning to bring into software development. They compare various aspects of software engineering (e.g., requirements, design, testing, and process) and work characteristics (e.g., skill variety, problem-solving and task identity) in both the ML systems and conventional software systems. 
Guo\cite{guo2019empirical} conducted a study on how various mainstream DL frameworks and platforms influence both DL software development and deployment in practice. 
Islam\cite{islam2019comprehensive}  studied 2716 high-quality posts from Stack Overflow and 500 bug fix commits from Github about five popular deep learning libraries to understand the types of bugs, root causes of bugs, impacts of bugs, bug-prone stage of deep learning pipeline as well as whether there are some common anti-patterns found in this buggy software. 
Najafabadi\cite{najafabadi2015deep} investigates some aspects of deep learning research that need further exploration to incorporate specific challenges introduced by big data analytics, including data of various formats and features. 
Fu et al.~\cite{fu2017easy} conducted a case study shows that applying a very simple optimizer, called differential evolution, to the fine-tune SVM can achieve similar results with much less time cost. This study casts doubts on the necessity of applying deep learning techniques in practice. 
Amershi et al.~\cite{amershi2019software} report their experience on developing AI-based systems in Microsoft. Their work identified three aspects of the AI domain that make it fundamentally different from prior software application.


Unfortunately, except for the empirical study papers, we did not find any papers present novel techniques and methods specifically for the software engineering phases of design, implementation, deployment, and maintenance. 


\subsection{Interviews}

To get deeper insights into designing the questionnaire, we conducted eight interviews with engineers from our industry partner companies, i.e., Baidu, Alibaba, and Huawei.
Because these companies have invested many resources in developing DL applications, DL frameworks, and related infrastructures, their engineers have sufficient experience and insights on our research topic.
Note that the goal of this study is to investigate the lacks and challenges in each phase of developing DL applications, thus, the interviewees consist of one project manager, one product designer, two testers, three developers, and one project maintainers.
We visited his/her company and conducted the interview face-to-face. 

During each interview, we kept to the following process: First, we explained to the interviewee the motivation of conducting the interview. 
And then we ask the interviewee to describe his/her job responsibility and discuss the challenges and lacks in building the DL application.
After that, we discussed the related research topics identified in the literature review with the interviewee.
We let the interviewee talk most of the time.
The whole interview lasts about 60 to 90 mins. 
We followed the methodology presented in~\cite{singer2014software,aniche2018modern} to decide when to stop interviewing, i.e. stopping interviews when the saturation of the themes is reached.

After we have finished the interview, each of the first three authors separately summarizes the lacks, challenges, and expectations mentioned in the interview. 
And then the three discuss each point to form a summary. 
Note that any discrepancy is discussed until a consensus is reached.
Finally, we send the summary back to the interviewee to confirm its correctness.

\subsection{Questionnaire Design}

Based on the literature review and interviews, we obtained a preliminary understanding of the lacks and challenges for the software engineering practice in the development of DL applications.
For each phase of developing DL applications, we summarized focuses of software researchers and our interviewees into Table~\ref{tab:questions}.
In Table~\ref{tab:questions}, the column \textit{Derived from} denotes the source of this foci come from, and the $|Questions|$ denotes the number of questions designed for this foci.
In total, we design 18 multiple choices questions based on these focuses.


\begin{table*}[!ht]
\caption{The focuses for each phase in the Questionnaire}
\label{tab:questions}
\begin{tabular}{|c|l|c|l|}
\hline
\textit{\textbf{Phase}}                                           & \multicolumn{1}{c|}{\textit{\textbf{Focuses}}}                                                & \textit{\textbf{|Question|}}              &  \textit{\textbf{Derived from}} \\ \hline
\multicolumn{1}{|c|}{\textit{\textbf{Resource Management}}}       & What are the motivations of redesigning and retraining the DNN models?      &    2                     &  \cite{fu2017easy},Interviews             \\ \hline
\textit{\textbf{Requirement Analysis}}                            & What are the motivations of employing DL techniques?                        &    2                   &  \cite{belani2019requirements}; Interviews                           \\ \hline
\multirow{1}{*}{\textit{\textbf{Design}}}                         & What are the challenges in the design of DL apps?                           &    3                   &  \cite{zhangempirical,amershi2019software,guo2019empirical}                                  \\ \hline
\multirow{3}{*}{\textit{\textbf{Implementation}}}                 & How to implement the DNN models?                                            &    1                   &  \cite{zhangempirical,zhang2019machine}                       \\ \cline{2-4} 
                                                & How to mitigate problems in the data-driven programming paradigm?           &    2                   &  \cite{zhangempirical,wan2019does}   \\ \cline{2-4} 
                                                & What are the primary training/testing data sources?                         &    1                    &  Interviews                        \\ \hline
\multirow{3}{*}{\textit{\textbf{Testing \& Debugging}}}           & What are the testing \& debugging methods ?                                 &    4                     &  \cite{tian2018deeptest,pei2017deepxplore,xie2019deephunter,guo2018dlfuzz,du2019deepstellar}        \\ \cline{2-4} 
                                                & What the metrics are used to guide the testing process?                     &    1                        &  \cite{ma2019deepct,ma2018deepgauge,ma2018deepmutation,sun2018testing}             \\  \hline
\multicolumn{1}{|c|}{\textit{\textbf{Deployment}}}                & How to reduce the size of DNN models?                                       &    1                    &  Interviews              \\ \hline
\multicolumn{1}{|c|}{\textit{\textbf{Maintenance}}}               & What are the motivations of redesigning and retraining the DNN models?      &    1                        &  Interviews             \\ \hline
\end{tabular}
\end{table*}

\section{Results and Findings}

\subsection{Demographics}

We include a number of  demographic questions in the questionnaire.
The demographic questions are designed to understand the background and experience of respondents.

Among all the respondents, 30 of them are junior practitioners with less than one-year work experience, 41 of them work in software engineering for 1-3 years, and 124 of them are experienced practitioners with more than 3 years of work experience.
Since experienced respondents are more likely to be practitioners of best practices in their corresponding field, intuitively our survey results of these practitioners reflect industry practices.

Meanwhile, we investigate the job roles of our respondents.
We have respondents who work as requirement engineers, software architects, software developers, software testers, and software operations engineers.
These practitioners perform tasks related to DL app development in phase of the software development life cycle corresponding to their job role.
Thus they can provide us real feedback from the industry.


We design a general question to figure out the primary influence brought by DL.
Difficulties in software engineering practices cause an increase in labor work.
8 primary tasks performed in software engineering are on the list.
Respondents were expected to choose the tasks where more labor work is required according to their own work experience.
To figure out which tasks pain points and difficulties exist, we divide respondents into groups according to their roles in software practice. The result is shown is Table \ref{tab:dlmorework}.

\begin{table*}[htbp]
\caption{Statistics on extra labor work required for tasks}
\begin{tabular}{|m{3cm}<{\centering}|m{1.3cm}<{\centering}|m{1.3cm}<{\centering}|m{1.3cm}<{\centering}|m{1.3cm}<{\centering}|m{1.3cm}<{\centering}|m{1.8cm}<{\centering}|m{1.8cm}<{\centering}|m{1.5cm}<{\centering}|} 
\hline
\textit{}                                                                                     & \footnotesize{\textit{\textbf{Problem definition}}} & \footnotesize{\textit{\textbf{Feasibility study}}} & \footnotesize{\textit{\textbf{Requirement analysis}}} & \footnotesize{\textit{\textbf{Summary design}}} & \footnotesize{\textit{\textbf{Detailed design}}} & \footnotesize{\textit{\textbf{Implementation \& unit test}}} & \footnotesize{\textit{\textbf{Test (integration/acceptance)}}} & \footnotesize{\textit{\textbf{Software maintenance}}} \\ \hline
\textit{\textbf{Requirement engineer}}                                                        & 42.86\%                              & 14.29\%                             & 35.71\%                                & 35.71\%                          & 21.43\%                           & 0.00\%                                & 7.14\%                                         & 7.14\%                                  \\ \hline
\textit{\textbf{Software architect}}                                                          & 64.00\%                        & 32.00\%                                & 60.00\%                          & 20.00\%                             & 12.00\%                              & 16.00\%                                  & 16.00\%                                           & 16.00\%                                    \\ \hline
\textit{\textbf{Developer}}                                                                   & 23.26\%                              & 39.53\%                             & 48.84\%                                & 11.63\%                          & 27.91\%                           & 18.60\%                               & 16.28\%                                        & 13.95\%                                 \\ \hline
\textit{\textbf{Tester}}                                                                      & 25.00\%                                 & 25.00\%                                & 35.23\%                                & 9.09\%                           & 20.45\%                           & 39.77\%                               & 54.55\%                               & 13.64\%                                 \\ \hline
\textit{\textbf{\begin{tabular}[l]{@{}c@{}} \small{Operation \&}\\\small{maintenance engineer}\end{tabular}}} & 25.00\%                                 & 50.00\%                                & 25.00\%                                   & 25.00\%                             & 50.00\%                              & 25.00\%                                  & 50.00\%                                  & 25.00\%                                    \\ \hline
$Total_{avg}$                                                                       & 30.61\%                             & 29.08\%                            & 40.82\%                               & 13.78\%                         & 20.41\%                          & 27.55                                 & 36.22\%                                       & 15.82\%                                 \\ \hline
\end{tabular}
\label{tab:dlmorework}
\end{table*}

In summary, requirement analysis, integration and acceptance testing, and problem definition are more likely to be labor-consuming.
We detail the summarization results for the designed questions regarding each phase in the software development life cycle. 
We further present the findings and provide answers to the research questions in the previous section.

\subsection{Requirement analysis}
Software engineers are expected to transform problems into reasoning logic reflected by the software system during requirement analysis.
However, it is well-recognized by respondents that requirement analysis is more difficult in DL applications.

While applications are considered to be more intelligent, fewer business rules are pre-defined.
Applications are expected to learn these rules from the given data.
In consequence, the reasoning logic is hidden behind, which makes it hard to be clarified by requirement engineers.
To make the case worse, the DNN does not take raw data as inputs, data needs to be prepared to form feature vectors.
However, despite the fact that the performance of the model depends highly on these feature vectors, what is learned by the DNN model is hard to interpret.
As a result, it is more difficult for requirement engineers to transfer the problem definition into specifications.

\begin{center}
\fbox{
  \parbox{0.46 \textwidth}{
      \textbf{Finding 1:} It is a challenging task to identify features over a large amount of data and verify its rationality in the requirement analysis phase.
  }
}
\end{center}

Difficult as it could be, according to our survey, practitioners from various industrial fields claim to provide intelligent services in their applications nowadays.
According to the result provided by respondents regarding the question "What are the motivation of employing DL techniques", there are two main motivations for practitioners to leverage this technique, as is shown in Table \ref{tab:reasons}. 

\begin{table}[htbp]
\caption{Motivations of employing DL techniques}
\begin{tabular}{@{}clc@{}}
\toprule
\textit{\textbf{Option}} & \multicolumn{1}{c}{\textit{\textbf{Reasons}}}             & \textit{\textbf{Votes(Ratios)}} \\ \midrule
\textit{A}               & \textit{\begin{tabular}[c]{@{}l@{}}DL is a promising feature needed to \\ be introduced as soon as possible.\end{tabular}}                                     & \textit{21.43\%}       \\ \midrule
\textit{B}               & \textit{\begin{tabular}[c]{@{}l@{}}DL technique has been widely used \\ in this area.\end{tabular}}                                                                                                                             & \textit{14.29\%}        \\ \midrule
\textit{C}               & \textit{\begin{tabular}[c]{@{}l@{}}Some preliminary research results \\ have been obtained via DL.\end{tabular}}                                               & \textit{35.71\%}       \\ \midrule
\textit{D}               & \textit{\begin{tabular}[c]{@{}l@{}}The feature to be implemented lacks \\ the defined rule definition but has \\ a large amount of business data\end{tabular}} & \textit{28.57\%}       \\ \bottomrule
\end{tabular}
\label{tab:reasons}
\end{table}

We find that about 14.29\% of respondents claimed that \textit{"DL technique has been widely used in this area"} even though it just becomes a hot topic in recent years.
We further filtered out responses submitted by them to check their experience and found that 60.71\% of them are experienced practitioners, 17.86\% of them have 1-3 years work experience while 21.43\% of them have work experience less than one year.

Meanwhile, we count the category of applications they worked on.
Shopping(39.29\%) is in the first place, where accurate and attractive recommendations are used to broker a deal, indicating that the DL technique performs remarkably in recommender systems.

\begin{center}
\fbox{%
  \parbox{0.46 \textwidth}{%
      \textbf{Finding 2:} Shopping is the leading category of applications where DNN models are used. DL techniques are widely adopted and achieves great performance in recommending commodities.
  }
}
\end{center}

Respondents who claimed that \textit{"DL is a promising feature needed to be introduced as soon as possible"} are considered to be normal developers attracted by technology.
About 21.43\% of practitioners are optimistic about DL techniques, which indicates that only about 20\% of developers in the market are newly attracted by DL.

Respondents who claimed that \textit{"Some preliminary research results have been obtained via DL"} were considered to be forerunners of software application development in their corresponding fields as they were more likely to try innovative technologies.
Owing to the breakthroughs DNNs achieved in some areas in recent years,  about 35.71\% of practitioners were willing to explore the possibility and performance of integrating DL into the current application.

Meanwhile, those who claimed that \textit{"The feature to be implemented lacks the defined rule definition but has a large amount of business data"} were considered to be practitioners in need of machine learning approaches.
About 28.57\% of practitioners regarded DL as a choice to mine value of business data since it is not easy to identify the relationships behind these data with human efforts.

\subsection{System design}
System design is the process of designing the elements of a system, including modules and components.
Design is prepared from the requirement specifications produced in the first phase.
We analyze on challenges in the design of DL applications in Table \ref{tab:systemdesign}.

\begin{table}[htbp]
\caption{Challenges in the design of DL applications}
\begin{tabular}{cccc}
\toprule
\textit{\textbf{Work experience}}                                                                         & \textit{\textbf{\textless 1 year}} & \textit{\textbf{1-3 years}} & \textit{\textbf{\textgreater 3 years}} \\ \hline
\textit{\textbf{Network model design}}                                                                    & 43.33\%                            & 64.29\%                     & 68.00\%                                \\ \hline
\textit{\textbf{Concurrency processing}}                                                                  & 46.67\%                            & 52.38\%                     & 53.60\%                                \\ \hline
\textit{\textbf{Traffic control}}                                                                         & 40.00\%                            & 28.57\%                     & 20.80\%                                \\ \hline
\textit{\textbf{\begin{tabular}[c]{@{}c@{}}No difference with \\ conventional applications\end{tabular}}} & 26.67\%                            & 26.19\%                     & 8.80\%                                 \\ 
\bottomrule
\end{tabular}
\label{tab:systemdesign}
\end{table}

Compared to practitioners who had less than 3 years of experience, only 8.80\% of the experienced practitioners claimed that there is no difference with conventional applications.
However, over 26.00\% of practitioners with less than 3 years of experience claimed that there is no difference between DL applications and conventional applications.
The ratio of people who think there is no difference drops from about 26.00\% to 8.80\%, indicating that it generally takes about 3 years for practitioners to understand the DL technique.
It is well-accepted that differences exist in the designing phase by them.

\begin{center}
\fbox{%
  \parbox{0.46 \textwidth}{%
      \textbf{Finding 3:} It takes about 3 years on average for a practitioner to become experienced guy in DL application development.
  }
}
\end{center}

According to our research, 63.45\% of the respondents on average held the opinion that the design of neural networks is a nodus to develop a DL app, even though 68.00\% of them had more than 3-year experience in software engineering.
More specifically, even with a series of open-source deep learning framework in the community, it is still very difficult to design a neural network which can perfectly fit the problem at hand.
To figure out the situation regarding practitioners with different experiences, we further divide the respondents into groups.

Concurrency processing is the second problem practitioners paid attention to developing DL applications. 
52.28\% of the respondents on average claimed about the concurrency processing problem. 
Similarly, experienced practitioners showed more concern on this problem since the ratio grows gradually from 46.67\% for junior practitioners, to 52.38\% for practitioners with 1-3 years of experience, to 53.6\% for experienced practitioners. 
In fact, DNN is not capable of handling requests in parallel, which can add to the time cost in case of a large amount of request simultaneously. 

\begin{center}
\fbox{%
  \parbox{0.46 \textwidth}{%
      \textbf{Finding 4:} Neural network model design and concurrent processing are problems to be solved in DL application development.
  }
}
\end{center}

Owing to the fact that DNNs in DL applications are trained over a huge amount of data.
The application should be capable of transfer more data when handling requests from application users.
About 25.38\% of practitioners on average showed their concern on network traffic control.
However, it is interesting that the ratio drops gradually, from 40.00\% for practitioners with less than 1-year experience, to 28.57\% for practitioners with 1-3 years experience, and finally to 20.80\% for practitioners with more than 3-year experience.
The result shows that on one hand, experienced practitioners are more capable of solving problems related to network traffic control.
On the other hand, there might be a shift of the focus practitioners work on from network traffic control to neural network design and concurrency processing.

Furthermore, we investigate on reasons why it is difficult to design a DNN.
The result is shown in Table \ref{tab:designdnn}.

\begin{table}[htbp]
\caption{Difficulties in designing deep neural networks}
\begin{tabular}{ccccc}
\toprule
\textit{\textbf{Work experience}}                                                                 & \textit{\textbf{\textless 1 year}} & \textit{\textbf{1-3 years}} & \textit{\textbf{\textgreater 3 years}} & \textit{\textbf{$Total_{avg}$}} \\ \hline
\textit{\textbf{Structure}}        & 36.67\%                            & 42.86\%                     & 61.60\%                                 & 53.81\%                 \\ \hline
\textit{\textbf{Performance}}      & 50.00\%                               & 69.05\%                     & 60.00\%                                   & 60.41\%                 \\ \hline
\textit{\textbf{Environment}} & 43.33\%                            & 54.76\%                     & 44.80\%                                 & 46.70\%                  \\ \hline
\textit{\textbf{Implementation}}   & 16.67\%                            & 33.33\%                     & 23.20\%                                 & 24.37\%                 \\ \bottomrule
\end{tabular}
\label{tab:designdnn}
\end{table}

About 60.41\% of practitioners on average claimed that the performance of the designed DNN model is unclear during the designing phase, indicating that the performance measurement of a DNN model remains a big problem in the designing phase. 
Meanwhile, 53.81\% of practitioners on average claimed that the structural details of the neural network model are not clear. 
Experienced practitioners are more likely to show their concerns on this problem since 61.60\% of them chose it.
Thus, how to reveal structure-related information of DNN is another problem to be solved. 
Besides, 47.60\% of practitioners on average claimed that the application environment is not clear. 

\begin{center}
\fbox{%
  \parbox{0.46 \textwidth}{%
      \textbf{Finding 5:} Performance measurement, structure design and application environment are three main factors that bring difficulties in designing DNNs.
  }
}
\end{center}

\subsection{Implementation}
Coding implementation is the main focus for developers.
In this phase, code blocks are implemented to work together to solve a problem in the real world.

A wide range of DL frameworks are available for building DNN models.
The popularity of each framework among industrial practitioners is shown in Table \ref{tab:frameworks}.
In total, TensorFlow(TF), PyTorch, and Caffe are the top 3 most popular deep learning frameworks. 
To figure out how practitioners implement the DNN models, according to the survey,  65.99\% of DL application practitioners on average use TensorFlow to build their neural network models. 30.96\% of them use PyTorch and 18.27\% of them use Caffe.

\begin{table}[htbp]
\caption{DL frameworks used to implement DNN models}
\begin{tabular}{ccccc}
\toprule
\textit{\textbf{Work experience}} & \textit{\textbf{\textless 1 year}} & \textit{\textbf{1-3 years}} & \textit{\textbf{\textgreater 3 years}} & \textit{\textbf{Total}} \\ \hline
\textit{\textbf{TensorFlow}}      & 60.00\%                            & 71.43\%                     & 65.60\%                                 & 65.99\%                 \\ \hline
\textit{\textbf{Caffe}}           & 13.33\%                            & 26.19\%                     & 16.80\%                                & 18.27\%                 \\ \hline
\textit{\textbf{PyTorch}}         & 13.33\%                            & 33.33\%                     & 34.40\%                                & 30.96\%                 \\ \hline
\textit{\textbf{Theano}}          & 13.33\%                            & 11.90\%                     & 10.40\%                                & 11.17\%                 \\ \hline
\textit{\textbf{PaddlePaddle}}    & 20.00\%                            & 14.29\%                     & 15.20\%                                & 15.74\%                 \\ \hline
\textit{\textbf{CNTK}}            & 6.67\%                             & 14.29\%                     & 11.20\%                                & 11.17\%                 \\ \hline
\textit{\textbf{MindSpore}}       & 6.67\%                             & 9.52\%                      & 13.60\%                                 & 11.68\%                 \\ \bottomrule
\end{tabular}
\label{tab:frameworks}
\end{table}

We further analyze the popularity of these DL frameworks among practitioners in groups according to their work experience.
For junior practitioners those who have less than 1 year work experience, about 60\% of them work with TensorFlow.
20\% of them work with PaddlePaddle, a deep learning framework released by Baidu.
For practitioners with 1-3 years work experience, 71.43\% of them work with TensorFlow, 33.33\% of them work with PyTorch, and 26.19\% of them work with Caffe.
For experienced practitioners with more than 3 years experience, 65.60\% of them work with TensorFlow, and 34.40\% of them work with PyTorch.

Since November, 2015 when the first public version of TensorFlow was released, it has won its popularity among practitioners with different work experiences.
PyTorch is more popular among experienced practitioners than among practitioners with less experience, indicating that it is not easy for junior practitioners to get familiar with this framework.

Another interesting thing we find is that, although PaddlePaddle(PP) was not in the list of most popular DL frameworks, it takes the second place in popularity among junior practitioners of DL applications.
For junior practitioners, detailed documentation and well-formed community can be a determinant in making a choice.

\begin{center}
\fbox{%
  \parbox{0.46 \textwidth}{%
      \textbf{Finding 6:} Most practitioners build their DL applications with the help of DL frameworks like TensorFlow and PyTorch.
  }
}
\end{center}

Most of the frameworks are open-source at the moment. 
However, MindSpore is a DL framework released by Huawei in August, 2019, which is not open-source yet. 
Thus those respondents who worked with MindSpore are expected to be practitioners from teams inside the company.
We filter out answers from these respondents to find out efforts that are put on other DL frameworks from competitors to implement MindSpore. 
Table \ref{tab:mindspore} shows the results. 

\begin{table}[htbp]
\caption{Frameworks that MindSpore practitioners use}
\begin{tabular}{cccccc}
\toprule
\textit{\textbf{TF}} & \textit{\textbf{Caffe}} & \textit{\textbf{PyTorch}} & \textit{\textbf{Theano}} & \textit{\textbf{PP}} & \textit{\textbf{CNTK}} \\ \hline
56.52\%                      & 47.83\%                 & 39.13\%                   & 34.78\%                  & 34.78\%                        & 43.48\%                \\ \bottomrule
\end{tabular}
\label{tab:mindspore}
\end{table}

On average, more than 30\% those respondents participant in building applications with DL frameworks released by other companies.
More efforts are put for popular DL frameworks, e.g. TensorFlow, Caffe, etc.
Another interesting thing we find is that about 43.48\% of those respondents work with CNTK, which is a less popular framework.
However, it is developed by Microsoft which is definitely a strong competitor in this area.

After some statistical analysis on DL frameworks adopted in the application development, we further investigate on practices done by practitioners.
Regarding conventional application development, implementation is an error-prune phase due to misuse of API, spelling error, uncaught exceptions, etc.
However, in DL application development, the model is expected to learn from data without any human efforts after it is implemented.
To avoid potential error introduced to the application during the implementation phase by this data-driven characteristic, developers works on methods to improve the correctness of both the application and the model.
The result is shown in Table \ref{tab:coding}.

\begin{table}[htbp]
\caption{Mitigate problems in the data-driven programming paradigm}
\begin{tabular}{ccccc}
\toprule
\textit{\textbf{Work experience}}                                                    & \textit{\textbf{\textless 1 year}} & \textit{\textbf{1-3 years}} & \textit{\textbf{\textgreater 3 years}} & \textit{\textbf{Total}} \\ \hline
\textit{\textbf{Less DL}}                                                            & 13.33\%                            & 19.05\%                     & 16.80\%                                 & 16.75\%                 \\ \hline
\textit{\textbf{QC of data}}                                               & 56.67\%                            & 59.52\%                     & 69.60\%                                 & 65.48\%                 \\ \hline
\textit{\textbf{\begin{tabular}[c]{@{}c@{}}Environment \\ consistency\end{tabular}}} & 60.00\%                               & 50.00\%                        & 56.80\%                                 & 55.84\%                 \\ \bottomrule
\end{tabular}
\label{tab:coding}
\vspace{-3mm}
\end{table}

On average, more than 65.48\% of practitioners try to solve the problem by quality control(QC) of data.
According to the result, experienced practitioners pay even more attention to the quality of the data than junior practitioners, indicating that it is effective to reduce errors in DL applications.
Though quality control of data is not necessary for developing conventional applications, developers regard it as an essential approach to control the quality of DL applications.

\begin{center}
\fbox{%
  \parbox{0.46 \textwidth}{%
      \textbf{Finding 7:} Quality control of data is an effective approach to solve problems brought by the data driven characteristic.
  }
}
\end{center}

Meanwhile, 55.84\% of practitioners on average regard environment consistency as a valid approach to avoid problems introduced by data driven.
Keeping the consistency of developing, testing and production environment is indeed a compromise to avoid triggering numerical related issues brought by environment differences.
However, practitioners are not offered with better alternatives to solve the problem at the moment.

\begin{center}
\fbox{%
  \parbox{0.46 \textwidth}{%
      \textbf{Finding 8:} Many developers try to avoid numerical related problems by keeping development environment, testing environment and production environment the same.
  }
}
\end{center}

In conventional applications, developers run unit test to ensure the correctness of individual functions.
To ensure the correctness of a developed DL model in the implementation phase, practices done by them are shown in Table \ref{tab:correctness}.

\begin{table}[htbp]
\caption{Approaches to ensure correctness in DL applications}
\begin{tabular}{ccccc}
\toprule
\textit{\textbf{Work experience}}        & \textit{\textbf{\textless 1 year}} & \textit{\textbf{1-3 years}} & \textit{\textbf{\textgreater 3 years}} & \textit{\textbf{Total}} \\ \hline
\textit{\textbf{Benchmark}} & 66.67\%                            & 52.38\%                     & 70.4\%                                 & 65.99\%                 \\ \hline
\textit{\textbf{Multiple models}}        & 20.00\%                            & 47.62\%                     & 51.2\%                                 & 45.69\%                 \\ \hline
\textit{\textbf{Code review}}            & 30.00                              & 28.57\%                     & 31.2\%                                 & 30.46\%                 \\ \hline
\textit{\textbf{Unit test}}              & 30.00\%                            & 16.67\%                     & 27.2\%                                 & 25.38\%                 \\ \bottomrule
\end{tabular}
\label{tab:correctness}
\end{table}

Unit test still takes part in guarantee the quality of code in the implementation phase.
However, to ensure the correctness of a DL application, 65.99\% of practitioners on average rely on model evaluation against benchmarks or large self-owned datasets. 
Meanwhile, 45.69\% of practitioners choose to implement multiple models to function together to ensure the correctness.
The ratio of this choice keeps growing considering the experience of practitioners.

\begin{center}
\fbox{%
  \parbox{0.46 \textwidth}{%
      \textbf{Finding 9:} Evaluation against benchmarks and implementing multiple models for a function are two main practices followed to provide correctness of the DL application in the implementation phase.
  }
}
\end{center}

\subsection{Testing}
Software testing is a process to evaluate the functionality with an intent to find whether the developed software meets the requirement or not and to provide guarantee for the quality of the software.
According to the result in Table \ref{tab:dlmorework}, this phase is considered to be one of the phase which receives a great impact after introducing the DL technique to application development.
The biggest difference between testing a conventional application and a DL application lies in that it requires additional work to  test whether the application under test is equipped with the specific knowledge.

When requirement analysis is done, testers begin to design test cases according to the requirement specifications.
Regarding conventional applications, the program is more interpretable as the processing logic is readable.
By analyzing the output together with various coverage criteria, the quality of such a software can be guaranteed.
However, owing to the fact that neural networks are multi-layer models with hyper-parameters used to learn features automatically, testers are not able to explain the true meaning of the numeric transformations inside at the moment.
In consequence, testers need to test whether the neural network model can make correct predictions in addition to testing the correctness, which adds to their labor work.

\begin{center}
\fbox{%
  \parbox{0.46 \textwidth}{%
      \textbf{Finding 10:} Testing whether the knowledge is obtained by a software remains a big problem in the testing phase.
  }
}
\end{center}

As a result, testers need to take a walk around to measure multiple metrics.
We analyze on the metrics testers focus on. 
The result is shown in Table \ref{tab:testingmetrics}.

\begin{table}[htbp]
\caption{Metrics to guide the testing process}
\vspace{-3mm}
\begin{tabular}{ccccc}
\toprule
\textit{\textbf{Work experience}} & \textit{\textbf{\textless 1 year}} & \textit{\textbf{1-3 years}} & \textit{\textbf{\textgreater 3 years}} & \textit{\textbf{Total}} \\ \hline
\textit{\textbf{Correctness}}     & 30.00\%                            & 50.00\%                     & 44.00\%                                & 43.15\%                 \\ \hline
\textit{\textbf{Performance}}     & 26.67\%                            & 57.14\%                     & 46.40\%                                & 45.69\%                 \\ \hline
\textit{\textbf{Compatibility}}   & 23.33\%                            & 45.24\%                     & 29.60\%                                & 31.98\%                 \\ \hline
\textit{\textbf{Robustness}}      & 56.67\%                            & 54.76\%                     & 57.60\%                                & 56.85\%                 \\ \bottomrule
\end{tabular}
\label{tab:testingmetrics}
\end{table}

Robustness is ranked as the most important metric to evaluate such an application.
Performance is another metric that testers focus on.
Due to the concurrency limitation brought by DNN models, performance bottleneck exists in handling a large amount of requests from the production environment.
Among these metrics, compatibility receives least attention.
However, application environment actually brings difficulties in the designing phase according to Finding 5.
We further investigate on the reason and find that it actually reacts to the environment consistency maintenance during development phase.
Because most of the testing environment is consistent with the developing environment, testers do not face with compatibility issues frequently during the testing phase at the moment.

We further conduct an analysis on testing practices to locate bugs once an error is triggered.
The result is shown in Table \ref{tab:locatebugs}.
\begin{table}[htbp]
\caption{Practices to locate bugs}
\vspace{-3mm}
\begin{tabular}{ccccc}
\toprule
\textit{\textbf{Work experience}}       & \textit{\textbf{\textless 1 year}} & \textit{\textbf{1-3 years}} & \textit{\textbf{\textgreater 3 years}} & \textit{\textbf{Total}} \\ \hline
\textit{\textbf{Bug locating tools}}    & 30.00\%                            & 35.71\%                     & 39.20\%                                & 37.06\%                 \\ \hline
\textit{\textbf{Self code review}} & 20.00\%                            & 23.81\%                     & 32.00\%                                & 28.43\%                 \\ \hline
\textit{\textbf{Cross code review}}     & 20.00\%                            & 35.71\%                     & 27.20\%                                & 27.92\%                 \\ \hline
\textit{\textbf{Retrain DNN}}           & 50.00\%                            & 33.33\%                     & 27.20\%                                & 31.98\%                 \\ \hline
\textit{\textbf{Log}}                   & 20.00\%                            & 28.57\%                     & 40.00\%                                & 34.52\%                 \\ \hline
\textit{\textbf{Break point}}            & 13.33\%                            & 28.57\%                     & 27.20\%                                & 25.38\%                 \\ \hline
\textit{\textbf{\textbf{\begin{tabular}[c]{@{}c@{}}Adversarial \\ samples\end{tabular}}}}   & 23.33\%                            & 38.10\%                     & 34.40\%                                & 33.5\%                  \\ \bottomrule
\end{tabular}
\label{tab:locatebugs}
\end{table}

According to the result, bug locating tools, log analysis, and adversarial samples are 3 main practices to locate bugs in DL applications.
Bug locating tools and log analysis are common practices done in testing conventional applications.
When testing DL applications, experienced practitioners show an appetite for these practices.
Besides, DNNs are found to be easily attacked by injecting small perturbations to the original input.
Various adversarial attack algorithms are proposed to detect defects of the model with the hope provide practitioners feedback about reasoning.

\begin{center}
\fbox{%
  \parbox{0.46 \textwidth}{%
      \textbf{Finding 11:} Bug locating tools, log analysis and adversarial sample are 3 main practices tester done to locate bugs in DL applications.
  }
}
\end{center}

We conduct another analysis on the challenges testers face with in testing and debugging DL applications.
Table \ref{tab:testingdifficulties} shows the result.

\begin{table}[htbp]
\caption{Challenges in testing \& debugging DL applications}
\vspace{-3mm}
\begin{tabular}{ccccc}
\toprule
\textit{\textbf{Work experience}}                                                             & \textit{\textbf{\textless 1 year}} & \textit{\textbf{1-3 years}} & \textit{\textbf{\textgreater 3 years}} & \textit{\textbf{Total}} \\ \hline
\textit{\textbf{Not enough data}}                                                             & 53.33\%                            & 66.67\%                     & 64.00\%                                & 62.94\%                 \\ \hline
\textit{\textbf{\begin{tabular}[c]{@{}c@{}}No assertion for\\ test data\end{tabular}}}        & 36.67\%                            & 26.19\%                     & 35.20\%                                & 33.50\%                 \\ \hline
\textit{\textbf{\begin{tabular}[c]{@{}c@{}}Lack of  system \\ level test case\end{tabular}}}  & 16.67\%                            & 42.86\%                     & 48.80\%                                & 42.64\%                 \\ \hline
\textit{\textbf{\begin{tabular}[c]{@{}c@{}}Low code statement\\ level coverage\end{tabular}}} & 16.67\%                            & 26.19\%                     & 14.40\%                                & 17.26\%                 \\ \bottomrule
\end{tabular}
\label{tab:testingdifficulties}
\end{table}

In testing conventional software applications, various code coverage metrics are adopted.
According to the survey, code statement level coverage is not difficult to reach.
However, testing still remains a big problem, which indicates that the result of coverage measurement based on code statement is not ideal in DL applications. 
Not getting enough testing data is the biggest problem they encounter in testing DL applications.
On the one hand, testing can only provide evidence to a buggy application instead of proving correctness for it. 
Thus a large amount of data is expected to achieve sufficient testing.
However, data collection is a high cost task due to the fact that some data is expensive and difficult to obtain in our daily life.
What's more, extra labor work is required to label all the testing data as oracle, which adds to the cost of the project.
On the other hand, testers did not know how to evaluate whether the knowledge learned by the DNN model is correct, they could only test it by adding as many samples as possible.

\begin{center}
\fbox{%
  \parbox{0.46 \textwidth}{%
      \textbf{Finding 12:} Due to the lack of testing data, it is difficult to test DL applications.
  }
}
\end{center}

In addition, it is claimed that existing test data are mainly inputs provided to the DNN model.
However, system level test cases are needed to check whether the whole system functions well.

Since there is an urgent requirement for test data, we analyze the primary data sources practitioners use in testing DL applications.
The result is shown in Table \ref{tab:testingsource}.

\begin{table}[htbp]
\caption{Primary testing data sources}
\vspace{-3mm}
\begin{tabular}{ccccc}
\toprule
\textit{\textbf{Work experience}}                                                            & \textit{\textbf{\textless 1 year}} & \textit{\textbf{1-3 years}} & \textit{\textbf{\textgreater 3 years}} & \textit{\textbf{Total}} \\ \hline
\textit{\textbf{Benchmark}}                                                                  & 53.33\%                            & 57.14\%                     & 56.00\%                                & 55.84\%                 \\ \hline
\textit{\textbf{\begin{tabular}[c]{@{}c@{}}Data obtained via \\ crowd-sourcing\end{tabular}}} & 23.33\%                            & 26.19\%                     & 34.40\%                                & 30.96\%                 \\ \hline
\textit{\textbf{\begin{tabular}[c]{@{}c@{}}Self-owned \\ business data\end{tabular}}}        & 46.67\%                            & 42.86\%                     & 52.00\%                                & 49.24\%                 \\ \hline
\textit{\textbf{\begin{tabular}[c]{@{}c@{}}Data \\ augmentation\end{tabular}}}               & 36.67\%                            & 52.38\%                     & 35.20\%                                & 39.09\%                 \\ \hline
\textit{\textbf{\begin{tabular}[c]{@{}c@{}}Manually \\ designed data\end{tabular}}}          & 26.67\%                            & 35.71\%                     & 41.60\%                                & 38.07\%                 \\ \bottomrule
\end{tabular}
\label{tab:testingsource}
\end{table}

55.84\% of practitioners leverage well-known benchmarks to test their models, followed by 49.24\% of them test DNN models with self-owned business data.
Meanwhile, practitioners try to obtain as much data as they can via multiple approaches including crowd-sourcing, data augmentation, etc.
It is notable that 41.60\% of experienced practitioners have to design data manually in their daily practice.

\subsection{Deployment and maintenance}
Once the application is ready for use, it will be deployed into the production environment.
Appeared and potential problems need to be solved from time to time later during the maintenance phase.

Despite the superior performance DNNs can achieve, models are space-consuming.
To find out practices practitioners done in deployment phase, we survey on approaches they follow in deploying models.
The result is shown in Table \ref{tab:deployment}.

\begin{table}[htbp]
\caption{Strategies to deploy industrial size DNN models}
\vspace{-3mm}
\begin{tabular}{ccccc}
\toprule
\textit{\textbf{Work experience}}                                                  & \textit{\textbf{\textless 1 year}} & \textit{\textbf{1-3 years}} & \textit{\textbf{\textgreater 3 years}} & \textit{\textbf{Total}} \\ \hline
\textit{\textbf{Model compression}}                                                & 11.54\%                            & 26.83\%                     & 24.19\%                                & 23.04\%                 \\ \hline
\textit{\textbf{\begin{tabular}[c]{@{}c@{}}Server side\\ deployment\end{tabular}}} & 61.54\%                            & 48.78\%                     & 51.61\%                                & 52.36\%                 \\ \hline
\textit{\textbf{Model pruning}}                                                    & 26.92\%                            & 24.39\%                     & 24.19\%                                & 24.61\%                 \\ \bottomrule
\end{tabular}
\label{tab:deployment}
\end{table}

Corresponding to the choice to keep consistency in development, testing and production environment, server side deployment is the first choice for 52.36\% of respondents.
Since resources are not that limited on servers as on embedded devices, server-side deployment allows practitioners to focus on performance of the model without considering complex external environments.
However, there are some practitioners try to decrease the size of the DNN model.
Model compression and model pruning are two common practices they perform to achieve the goal.

By deploying DNN models on the server side, practitioners are able to avoid some model fitness and compatibility problems.
Although it brings flexibility to maintain the model, challenges still exist in the maintenance of DL applications.
The result of the question "What are the challenges in the maintenance of DL apps", as shown in Table \ref{tab:maintain}.

\begin{table}[htbp]
\caption{Challenges in the maintenance of DL applications}
\begin{tabular}{ccccc}
\toprule
\textit{\textbf{Work experience}}                                                                            & \textit{\textbf{\textless 1 year}} & \textit{\textbf{1-3 years}} & \textit{\textbf{\textgreater 3 years}} & \textit{\textbf{Total}}     \\ \hline
\textit{\textbf{\begin{tabular}[c]{@{}c@{}}Multiple model\\ maintenance\end{tabular}}}                       & 25.93\%                            & 17.50\%                     & 15.45\%                                & 16.75\%                     \\ \hline
\textit{\textbf{\begin{tabular}[c]{@{}c@{}}Extra efforts to\\ evaluate and \\maintain QoD\end{tabular}}}                    & 29.63\%                            & 42.50\%                     & 34.15\%                                & 34.01\%                     \\ \hline
\textit{\textbf{\begin{tabular}[c]{@{}c@{}}Frequency increase\\ due to low\\ interpretability\end{tabular}}} & 40.74\%                            & 20.00\%                     & 21.95\%                                & 23.35\%                     \\ \hline
\textit{\textbf{Higher labor cost}}                                                                          & 3.70\%                             & 20.00\%                     & 28.46\%                                & \multicolumn{1}{l}{22.34\%} \\
\bottomrule
\end{tabular}
\label{tab:maintain}
\end{table}

Extra efforts to evaluate and maintain the quality of data is recognized as the primary difficulty in DL application maintenance by 34.01\% of respondents. 
Meanwhile, due to the fact that models need to be retrained from time to time to keep itself up-to-date with the knowledge evolvement, practitioners need to take the quality of data(QoD) into consideration as the application keeps receiving new business data which may contain new domain knowledge. 

\begin{center}
\fbox{%
  \parbox{0.46 \textwidth}{%
      \textbf{Finding 13:} Quality control of data remains a big problem in the phase of software maintenance.
  }
}
\end{center}

According to the result, multiple model maintenance does not actually add much to the difficulties in maintaining DL applications.
In contrast, the low interpretability of each individual model makes it difficult to maintain as why the error occurs cannot be explained.

\section{Discussion}

\subsection{Implications}

\textbf{For Practitioners:} 
Junior practitioners in software engineering are often confused about what practices to follow due to the new data-driven paradigm brought by DL. 
\begin{itemize} 
\item According to Finding 6, junior practitioners of DL application development is recommended to start with well-known deep learning frameworks. 
\item According to Finding 8, keeping DNN components of an application on the server-side, as well as keep consistency in development, testing and production environment, can protect practitioners from potential problems caused by differences in the environment. 
\item According to Finding 9, to improve the correctness and robustness of DL applications, practitioners can leverage benchmarks to provide sufficient test to the DL application. Meanwhile, implementing multiple models can avoid the problem to some extends. 
\end{itemize}

\noindent
\textbf{For Researchers:} 
Our findings also highlight opportunities for software engineering researchers to build tools and techniques that can help practitioners improve the quality of DL applications. 

\begin{itemize}
\item According to Finding 1, 7 and 13, data feature identification, quality control, and evaluation are challenging tasks in DL application development.  Researchers are encouraged to build tools to detect and highlight features from massive data, which can give requirement engineers heuristics about data preparation. 
Tools that help in quality control of data like dataset bias detection, distribution evaluation, etc. are also required to provide guarantees for training and testing datasets. 
\item According to Finding 4, concurrent processing is a primary factor that restricts the adoption of deep learning. Researches on approaches to enable concurrency in DNN can attract more practitioners to apply it in real applications. 
Together with Finding 5, tools to provide real-time evaluation and feedback for software architects can better guide them in designing more powerful models to make better use of the business data. 
\item According to Finding 11, bug locating tools and logs are of great help in locating bugs. 
In deep learning, multiple adversarial attack algorithms are proposed to expose potential defects inside the model. 
Researchers are expected to work on approaches and tools to map adversarial samples into the reasoning process to find out the location where the error is triggered, which can be of great help in further interpreting deep neural networks. 
\item According to Finding 12, practitioners are in lack of enough data to evaluate the  DL application under test. 
Since some data are hard from collect in our daily life, tools that can produce inputs that maintain the same semantics can further help engineers to analyze DNN models. 
Meanwhile, research on approaches to evaluating whether the test is sufficient can help practitioners find a balance between model testing and data collection.
\end{itemize}

\subsection{Threats to Validity}
\textbf{External Validity:}
Focusing on the topic of the development of DL applications, We summarized 13 relevant papers that are published in software engineering venues. 
Through this literature survey, we get numerous insights, yet it may be a small sample. 
Also, our interviewees only come from three big companies. 
To mitigate this threat, we survey 195 respondents from various small and large organizations with different backgrounds. 
Still, our findings may not be generalized to all practitioners. 

\noindent
\textbf{Internal Validity:} 
One of the primary internal validity comes from the completeness of our survey. 
Respondents may have different opinions beyond the ones we summarized in the questionnaire. 
To reduce bias in the survey, we keep all questions open-ended and let respondents express their opinions. 
Moreover, it is possible that some respondents do not understand the questions well, or the questions are beyond their work experience. 
To minimize this threat, for each question, we provide the option ``I don't know" to let respondents make the proper feedback. 
Another threat lays in the creation of questions, which are summarized by literature surveys and interviews with individuals. 
These interviews only reflect the perspectives of individuals, and thus may introduce bias in our study. 
However, each of the three authors vetted through the questions, as well as options, created by the other authors to mitigate this threat. 
Also, before we distribute the questionnaire, we send it to our industrial interviewees to validate each question and its options. 

\section{Related Work}

Empirical study in software engineering has been conducted for several decades and achieves significant recognition in the broader software engineering research community~\cite{seaman1999qualitative}. Rapid changes in competitive threats, stakeholder preferences, development technology, and time-to-market pressures make pre-specified requirements inappropriate. 
Card~\cite{card1986empirical} conducted an empirical study on software design theory in one specific environment by examining multiple metrics including module size, module strength, data coupling, unreferenced variables, etc. The result shows that some recommended design practices can be ineffective in this environment despite their intuitive appeal. 
To estimate the performance of software maintenance, Bankder\cite{banker1998software} conducted a field study to provides insight into how performance in software maintenance can be improved by improving the efficacy of design and development procedures.
To understand open software development practices, Scacchi\cite{scacchi2001software} conducted a comparative case study across open source communities.
Cao~\cite{cao2008agile} conducted an empirical analysis to figure out the requirement engineering practices that agile developers follow and the benefits and challenges these practices present.
Itkonen\cite{itkonen2009testers} presents a study on the manual testing practices in four software development companies and identified 22 manual testing practices and further compared it with traditional test practices.
To understand the benefits, risks, and limitations of using social media in software development, Storey\cite{storey2010impact} proposes and answers a set of pertinent research questions around community involvement, project coordination, and management, as well as individual software development activities.
Daka et al.~\cite{daka2014survey} surveyed 225 developers to understand unit testing practices such as motivation of developers, their usage of automation tools, and their challenges. 
Wan et al. ~\cite{wan2019does} performed a mixture of qualitative and quantitative studies with 14 interviewees and 342 questionnaire
respondents to investigate the impacts of machine learning on the software development.

Different from these work, in this study, we complement existing empirical studies by conducting a comprehensive survey with 195 industrial practitioners to understand the characteristics of each phase of the DL application development. We also invited respondents to provide their rationale for the two hottest topics, i.e., testing and debugging. These findings and feedback provide us with a comprehensive understanding of the vision and challenges of the DL application development. 


\section{Conclusion}
Even though deep learning is an efficient approach to deal with big data and to make apps more intelligent, challenges and lacks in practices of DL applications development are not clear.
In this paper, we investigate the challenges and lacks in practice when developing a DL application.
We interview our industry partner companies to find out the lacks and challenges in each phase of software development.
We further survey 195 practitioners of DL applications from different companies.
Our survey results indicate that the data-driven paradigm of deep learning brings challenges to each phase of the software development life cycle.
We conclude 13 findings from the results.
Based on these findings, we make a discussion and propose 7 actionable recommendations for DL application practitioners, as well as potential research directions for researchers to explore.
Progress in such directions would further promote the development of DNN as well as DL applications.

\bibliographystyle{ACM-Reference-Format}
\bibliography{ref}


\begin{thebibliography}{37}


\ifx \showCODEN    \undefined \def \showCODEN     #1{\unskip}     \fi
\ifx \showDOI      \undefined \def \showDOI       #1{#1}\fi
\ifx \showISBNx    \undefined \def \showISBNx     #1{\unskip}     \fi
\ifx \showISBNxiii \undefined \def \showISBNxiii  #1{\unskip}     \fi
\ifx \showISSN     \undefined \def \showISSN      #1{\unskip}     \fi
\ifx \showLCCN     \undefined \def \showLCCN      #1{\unskip}     \fi
\ifx \shownote     \undefined \def \shownote      #1{#1}          \fi
\ifx \showarticletitle \undefined \def \showarticletitle #1{#1}   \fi
\ifx \showURL      \undefined \def \showURL       {\relax}        \fi
\providecommand\bibfield[2]{#2}
\providecommand\bibinfo[2]{#2}
\providecommand\natexlab[1]{#1}
\providecommand\showeprint[2][]{arXiv:#2}

\bibitem[\protect\citeauthoryear{Amershi, Begel, Bird, DeLine, Gall, Kamar,
  Nagappan, Nushi, and Zimmermann}{Amershi et~al\mbox{.}}{2019}]%
        {amershi2019software}
\bibfield{author}{\bibinfo{person}{Saleema Amershi}, \bibinfo{person}{Andrew
  Begel}, \bibinfo{person}{Christian Bird}, \bibinfo{person}{Robert DeLine},
  \bibinfo{person}{Harald Gall}, \bibinfo{person}{Ece Kamar},
  \bibinfo{person}{Nachiappan Nagappan}, \bibinfo{person}{Besmira Nushi}, {and}
  \bibinfo{person}{Thomas Zimmermann}.} \bibinfo{year}{2019}\natexlab{}.
\newblock \showarticletitle{Software engineering for machine learning: a case
  study}. In \bibinfo{booktitle}{\emph{Proceedings of the 41st International
  Conference on Software Engineering: Software Engineering in Practice}}. IEEE
  Press, \bibinfo{pages}{291--300}.
\newblock


\bibitem[\protect\citeauthoryear{Aniche, Treude, Steinmacher, Wiese, Pinto,
  Storey, and Gerosa}{Aniche et~al\mbox{.}}{2018}]%
        {aniche2018modern}
\bibfield{author}{\bibinfo{person}{Maur{\'\i}cio Aniche},
  \bibinfo{person}{Christoph Treude}, \bibinfo{person}{Igor Steinmacher},
  \bibinfo{person}{Igor Wiese}, \bibinfo{person}{Gustavo Pinto},
  \bibinfo{person}{Margaret-Anne Storey}, {and}
  \bibinfo{person}{Marco~Aur{\'e}lio Gerosa}.} \bibinfo{year}{2018}\natexlab{}.
\newblock \showarticletitle{How modern news aggregators help development
  communities shape and share knowledge}. In \bibinfo{booktitle}{\emph{2018
  IEEE/ACM 40th International Conference on Software Engineering (ICSE)}}.
  IEEE, \bibinfo{pages}{499--510}.
\newblock


\bibitem[\protect\citeauthoryear{Banker, Davis, and Slaughter}{Banker
  et~al\mbox{.}}{1998}]%
        {banker1998software}
\bibfield{author}{\bibinfo{person}{Rajiv~D Banker}, \bibinfo{person}{Gordon~B
  Davis}, {and} \bibinfo{person}{Sandra~A Slaughter}.}
  \bibinfo{year}{1998}\natexlab{}.
\newblock \showarticletitle{Software development practices, software
  complexity, and software maintenance performance: A field study}.
\newblock \bibinfo{journal}{\emph{Management science}} \bibinfo{volume}{44},
  \bibinfo{number}{4} (\bibinfo{year}{1998}), \bibinfo{pages}{433--450}.
\newblock


\bibitem[\protect\citeauthoryear{Belani, Vukovi{\'c}, and Car}{Belani
  et~al\mbox{.}}{2019}]%
        {belani2019requirements}
\bibfield{author}{\bibinfo{person}{Hrvoje Belani}, \bibinfo{person}{Marin
  Vukovi{\'c}}, {and} \bibinfo{person}{{\v{Z}}eljka Car}.}
  \bibinfo{year}{2019}\natexlab{}.
\newblock \showarticletitle{Requirements Engineering Challenges in Building
  AI-Based Complex Systems}.
\newblock \bibinfo{journal}{\emph{arXiv preprint arXiv:1908.11791}}
  (\bibinfo{year}{2019}).
\newblock


\bibitem[\protect\citeauthoryear{Bojarski, Del~Testa, Dworakowski, Firner,
  Flepp, Goyal, Jackel, Monfort, Muller, and Zhang}{Bojarski
  et~al\mbox{.}}{2016}]%
        {bojarski_end_2016}
\bibfield{author}{\bibinfo{person}{Mariusz Bojarski}, \bibinfo{person}{Davide
  Del~Testa}, \bibinfo{person}{Daniel Dworakowski}, \bibinfo{person}{Bernhard
  Firner}, \bibinfo{person}{Beat Flepp}, \bibinfo{person}{Prasoon Goyal},
  \bibinfo{person}{Lawrence~D. Jackel}, \bibinfo{person}{Mathew Monfort},
  \bibinfo{person}{Urs Muller}, {and} \bibinfo{person}{Jiakai Zhang}.}
  \bibinfo{year}{2016}\natexlab{}.
\newblock \showarticletitle{End to end learning for self-driving cars}.
\newblock \bibinfo{journal}{\emph{arXiv preprint arXiv:1604.07316}}
  (\bibinfo{year}{2016}).
\newblock


\bibitem[\protect\citeauthoryear{Cao and Ramesh}{Cao and Ramesh}{2008}]%
        {cao2008agile}
\bibfield{author}{\bibinfo{person}{Lan Cao} {and}
  \bibinfo{person}{Balasubramaniam Ramesh}.} \bibinfo{year}{2008}\natexlab{}.
\newblock \showarticletitle{Agile requirements engineering practices: An
  empirical study}.
\newblock \bibinfo{journal}{\emph{IEEE software}} \bibinfo{volume}{25},
  \bibinfo{number}{1} (\bibinfo{year}{2008}), \bibinfo{pages}{60--67}.
\newblock


\bibitem[\protect\citeauthoryear{Card, Church, and Agresti}{Card
  et~al\mbox{.}}{1986}]%
        {card1986empirical}
\bibfield{author}{\bibinfo{person}{David~N Card}, \bibinfo{person}{Victor~E
  Church}, {and} \bibinfo{person}{William~W Agresti}.}
  \bibinfo{year}{1986}\natexlab{}.
\newblock \showarticletitle{An empirical study of software design practices}.
\newblock \bibinfo{journal}{\emph{IEEE Transactions on Software Engineering}}
  \bibinfo{number}{2} (\bibinfo{year}{1986}), \bibinfo{pages}{264--271}.
\newblock


\bibitem[\protect\citeauthoryear{Daka and Fraser}{Daka and Fraser}{2014}]%
        {daka2014survey}
\bibfield{author}{\bibinfo{person}{Ermira Daka} {and} \bibinfo{person}{Gordon
  Fraser}.} \bibinfo{year}{2014}\natexlab{}.
\newblock \showarticletitle{A survey on unit testing practices and problems}.
  In \bibinfo{booktitle}{\emph{2014 IEEE 25th International Symposium on
  Software Reliability Engineering}}. IEEE, \bibinfo{pages}{201--211}.
\newblock


\bibitem[\protect\citeauthoryear{Du, Xie, Li, Ma, Liu, and Zhao}{Du
  et~al\mbox{.}}{2019}]%
        {du2019deepstellar}
\bibfield{author}{\bibinfo{person}{Xiaoning Du}, \bibinfo{person}{Xiaofei Xie},
  \bibinfo{person}{Yi Li}, \bibinfo{person}{Lei Ma}, \bibinfo{person}{Yang
  Liu}, {and} \bibinfo{person}{Jianjun Zhao}.} \bibinfo{year}{2019}\natexlab{}.
\newblock \showarticletitle{Deepstellar: model-based quantitative analysis of
  stateful deep learning systems}. In \bibinfo{booktitle}{\emph{Proceedings of
  the 2019 27th ACM Joint Meeting on European Software Engineering Conference
  and Symposium on the Foundations of Software Engineering}}. ACM,
  \bibinfo{pages}{477--487}.
\newblock


\bibitem[\protect\citeauthoryear{Fu and Menzies}{Fu and Menzies}{2017}]%
        {fu2017easy}
\bibfield{author}{\bibinfo{person}{Wei Fu} {and} \bibinfo{person}{Tim
  Menzies}.} \bibinfo{year}{2017}\natexlab{}.
\newblock \showarticletitle{Easy over hard: A case study on deep learning}. In
  \bibinfo{booktitle}{\emph{Proceedings of the 2017 11th Joint Meeting on
  Foundations of Software Engineering}}. ACM, \bibinfo{pages}{49--60}.
\newblock


\bibitem[\protect\citeauthoryear{Gulshan, Peng, Coram, Stumpe, Wu,
  Narayanaswamy, Venugopalan, Widner, Madams, Cuadros, et~al\mbox{.}}{Gulshan
  et~al\mbox{.}}{2016}]%
        {gulshan2016development}
\bibfield{author}{\bibinfo{person}{Varun Gulshan}, \bibinfo{person}{Lily Peng},
  \bibinfo{person}{Marc Coram}, \bibinfo{person}{Martin~C Stumpe},
  \bibinfo{person}{Derek Wu}, \bibinfo{person}{Arunachalam Narayanaswamy},
  \bibinfo{person}{Subhashini Venugopalan}, \bibinfo{person}{Kasumi Widner},
  \bibinfo{person}{Tom Madams}, \bibinfo{person}{Jorge Cuadros},
  {et~al\mbox{.}}} \bibinfo{year}{2016}\natexlab{}.
\newblock \showarticletitle{Development and validation of a deep learning
  algorithm for detection of diabetic retinopathy in retinal fundus
  photographs}.
\newblock \bibinfo{journal}{\emph{Jama}} \bibinfo{volume}{316},
  \bibinfo{number}{22} (\bibinfo{year}{2016}), \bibinfo{pages}{2402--2410}.
\newblock


\bibitem[\protect\citeauthoryear{Guo, Jiang, Zhao, Chen, and Sun}{Guo
  et~al\mbox{.}}{2018}]%
        {guo2018dlfuzz}
\bibfield{author}{\bibinfo{person}{Jianmin Guo}, \bibinfo{person}{Yu Jiang},
  \bibinfo{person}{Yue Zhao}, \bibinfo{person}{Quan Chen}, {and}
  \bibinfo{person}{Jiaguang Sun}.} \bibinfo{year}{2018}\natexlab{}.
\newblock \showarticletitle{Dlfuzz: Differential fuzzing testing of deep
  learning systems}. In \bibinfo{booktitle}{\emph{Proceedings of the 2018 26th
  ACM Joint Meeting on European Software Engineering Conference and Symposium
  on the Foundations of Software Engineering}}. ACM, \bibinfo{pages}{739--743}.
\newblock


\bibitem[\protect\citeauthoryear{Guo, Chen, Xie, Ma, Hu, Liu, Liu, Zhao, and
  Li}{Guo et~al\mbox{.}}{2019}]%
        {guo2019empirical}
\bibfield{author}{\bibinfo{person}{Qianyu Guo}, \bibinfo{person}{Sen Chen},
  \bibinfo{person}{Xiaofei Xie}, \bibinfo{person}{Lei Ma},
  \bibinfo{person}{Qiang Hu}, \bibinfo{person}{Hongtao Liu},
  \bibinfo{person}{Yang Liu}, \bibinfo{person}{Jianjun Zhao}, {and}
  \bibinfo{person}{Xiaohong Li}.} \bibinfo{year}{2019}\natexlab{}.
\newblock \showarticletitle{An Empirical Study towards Characterizing Deep
  Learning Development and Deployment across Different Frameworks and
  Platforms}.
\newblock \bibinfo{journal}{\emph{arXiv preprint arXiv:1909.06727}}
  (\bibinfo{year}{2019}).
\newblock


\bibitem[\protect\citeauthoryear{Huval, Wang, Tandon, Kiske, Song,
  Pazhayampallil, Andriluka, Rajpurkar, Migimatsu, Cheng-Yue,
  et~al\mbox{.}}{Huval et~al\mbox{.}}{2015}]%
        {huval2015empirical}
\bibfield{author}{\bibinfo{person}{Brody Huval}, \bibinfo{person}{Tao Wang},
  \bibinfo{person}{Sameep Tandon}, \bibinfo{person}{Jeff Kiske},
  \bibinfo{person}{Will Song}, \bibinfo{person}{Joel Pazhayampallil},
  \bibinfo{person}{Mykhaylo Andriluka}, \bibinfo{person}{Pranav Rajpurkar},
  \bibinfo{person}{Toki Migimatsu}, \bibinfo{person}{Royce Cheng-Yue},
  {et~al\mbox{.}}} \bibinfo{year}{2015}\natexlab{}.
\newblock \showarticletitle{An empirical evaluation of deep learning on highway
  driving}.
\newblock \bibinfo{journal}{\emph{arXiv preprint arXiv:1504.01716}}
  (\bibinfo{year}{2015}).
\newblock


\bibitem[\protect\citeauthoryear{Islam, Nguyen, Pan, and Rajan}{Islam
  et~al\mbox{.}}{2019}]%
        {islam2019comprehensive}
\bibfield{author}{\bibinfo{person}{Md~Johirul Islam}, \bibinfo{person}{Giang
  Nguyen}, \bibinfo{person}{Rangeet Pan}, {and} \bibinfo{person}{Hridesh
  Rajan}.} \bibinfo{year}{2019}\natexlab{}.
\newblock \showarticletitle{A Comprehensive Study on Deep Learning Bug
  Characteristics}.
\newblock \bibinfo{journal}{\emph{arXiv preprint arXiv:1906.01388}}
  (\bibinfo{year}{2019}).
\newblock


\bibitem[\protect\citeauthoryear{Itkonen, Mantyla, and Lassenius}{Itkonen
  et~al\mbox{.}}{2009}]%
        {itkonen2009testers}
\bibfield{author}{\bibinfo{person}{Juha Itkonen}, \bibinfo{person}{Mika~V
  Mantyla}, {and} \bibinfo{person}{Casper Lassenius}.}
  \bibinfo{year}{2009}\natexlab{}.
\newblock \showarticletitle{How do testers do it? An exploratory study on
  manual testing practices}. In \bibinfo{booktitle}{\emph{2009 3rd
  International Symposium on Empirical Software Engineering and Measurement}}.
  IEEE, \bibinfo{pages}{494--497}.
\newblock


\bibitem[\protect\citeauthoryear{Katz, Barrett, Dill, Julian, and
  Kochenderfer}{Katz et~al\mbox{.}}{2017}]%
        {katz2017reluplex}
\bibfield{author}{\bibinfo{person}{Guy Katz}, \bibinfo{person}{Clark Barrett},
  \bibinfo{person}{David~L Dill}, \bibinfo{person}{Kyle Julian}, {and}
  \bibinfo{person}{Mykel~J Kochenderfer}.} \bibinfo{year}{2017}\natexlab{}.
\newblock \showarticletitle{Reluplex: An efficient SMT solver for verifying
  deep neural networks}. In \bibinfo{booktitle}{\emph{International Conference
  on Computer Aided Verification}}. Springer, \bibinfo{pages}{97--117}.
\newblock


\bibitem[\protect\citeauthoryear{Ma, Juefei-Xu, Xue, Li, Li, Liu, and Zhao}{Ma
  et~al\mbox{.}}{2019}]%
        {ma2019deepct}
\bibfield{author}{\bibinfo{person}{Lei Ma}, \bibinfo{person}{Felix Juefei-Xu},
  \bibinfo{person}{Minhui Xue}, \bibinfo{person}{Bo Li}, \bibinfo{person}{Li
  Li}, \bibinfo{person}{Yang Liu}, {and} \bibinfo{person}{Jianjun Zhao}.}
  \bibinfo{year}{2019}\natexlab{}.
\newblock \showarticletitle{DeepCT: Tomographic Combinatorial Testing for Deep
  Learning Systems}. In \bibinfo{booktitle}{\emph{2019 IEEE 26th International
  Conference on Software Analysis, Evolution and Reengineering (SANER)}}. IEEE,
  \bibinfo{pages}{614--618}.
\newblock


\bibitem[\protect\citeauthoryear{Ma, Juefei-Xu, Zhang, Sun, Xue, Li, Chen, Su,
  Li, Liu, et~al\mbox{.}}{Ma et~al\mbox{.}}{2018a}]%
        {ma2018deepgauge}
\bibfield{author}{\bibinfo{person}{Lei Ma}, \bibinfo{person}{Felix Juefei-Xu},
  \bibinfo{person}{Fuyuan Zhang}, \bibinfo{person}{Jiyuan Sun},
  \bibinfo{person}{Minhui Xue}, \bibinfo{person}{Bo Li},
  \bibinfo{person}{Chunyang Chen}, \bibinfo{person}{Ting Su},
  \bibinfo{person}{Li Li}, \bibinfo{person}{Yang Liu}, {et~al\mbox{.}}}
  \bibinfo{year}{2018}\natexlab{a}.
\newblock \showarticletitle{Deepgauge: Multi-granularity testing criteria for
  deep learning systems}. In \bibinfo{booktitle}{\emph{Proceedings of the 33rd
  ACM/IEEE International Conference on Automated Software Engineering}}. ACM,
  \bibinfo{pages}{120--131}.
\newblock


\bibitem[\protect\citeauthoryear{Ma, Zhang, Sun, Xue, Li, Juefei-Xu, Xie, Li,
  Liu, Zhao, et~al\mbox{.}}{Ma et~al\mbox{.}}{2018b}]%
        {ma2018deepmutation}
\bibfield{author}{\bibinfo{person}{Lei Ma}, \bibinfo{person}{Fuyuan Zhang},
  \bibinfo{person}{Jiyuan Sun}, \bibinfo{person}{Minhui Xue},
  \bibinfo{person}{Bo Li}, \bibinfo{person}{Felix Juefei-Xu},
  \bibinfo{person}{Chao Xie}, \bibinfo{person}{Li Li}, \bibinfo{person}{Yang
  Liu}, \bibinfo{person}{Jianjun Zhao}, {et~al\mbox{.}}}
  \bibinfo{year}{2018}\natexlab{b}.
\newblock \showarticletitle{Deepmutation: Mutation testing of deep learning
  systems}. In \bibinfo{booktitle}{\emph{2018 IEEE 29th International Symposium
  on Software Reliability Engineering (ISSRE)}}. IEEE,
  \bibinfo{pages}{100--111}.
\newblock


\bibitem[\protect\citeauthoryear{Najafabadi, Villanustre, Khoshgoftaar, Seliya,
  Wald, and Muharemagic}{Najafabadi et~al\mbox{.}}{2015}]%
        {najafabadi2015deep}
\bibfield{author}{\bibinfo{person}{Maryam~M Najafabadi},
  \bibinfo{person}{Flavio Villanustre}, \bibinfo{person}{Taghi~M Khoshgoftaar},
  \bibinfo{person}{Naeem Seliya}, \bibinfo{person}{Randall Wald}, {and}
  \bibinfo{person}{Edin Muharemagic}.} \bibinfo{year}{2015}\natexlab{}.
\newblock \showarticletitle{Deep learning applications and challenges in big
  data analytics}.
\newblock \bibinfo{journal}{\emph{Journal of Big Data}} \bibinfo{volume}{2},
  \bibinfo{number}{1} (\bibinfo{year}{2015}), \bibinfo{pages}{1}.
\newblock


\bibitem[\protect\citeauthoryear{Pei, Cao, Yang, and Jana}{Pei
  et~al\mbox{.}}{2017}]%
        {pei2017deepxplore}
\bibfield{author}{\bibinfo{person}{Kexin Pei}, \bibinfo{person}{Yinzhi Cao},
  \bibinfo{person}{Junfeng Yang}, {and} \bibinfo{person}{Suman Jana}.}
  \bibinfo{year}{2017}\natexlab{}.
\newblock \showarticletitle{Deepxplore: Automated whitebox testing of deep
  learning systems}. In \bibinfo{booktitle}{\emph{Proceedings of the 26th
  Symposium on Operating Systems Principles}}. ACM, \bibinfo{pages}{1--18}.
\newblock


\bibitem[\protect\citeauthoryear{Scacchi}{Scacchi}{2001}]%
        {scacchi2001software}
\bibfield{author}{\bibinfo{person}{Walt Scacchi}.}
  \bibinfo{year}{2001}\natexlab{}.
\newblock \showarticletitle{Software development practices in open software
  development communities: a comparative case study}. In
  \bibinfo{booktitle}{\emph{Proceedings of 1st Workshop on Open Source Software
  Engineering}}.
\newblock


\bibitem[\protect\citeauthoryear{Seaman}{Seaman}{1999}]%
        {seaman1999qualitative}
\bibfield{author}{\bibinfo{person}{Carolyn~B. Seaman}.}
  \bibinfo{year}{1999}\natexlab{}.
\newblock \showarticletitle{Qualitative methods in empirical studies of
  software engineering}.
\newblock \bibinfo{journal}{\emph{IEEE Transactions on software engineering}}
  \bibinfo{volume}{25}, \bibinfo{number}{4} (\bibinfo{year}{1999}),
  \bibinfo{pages}{557--572}.
\newblock


\bibitem[\protect\citeauthoryear{Singer, Figueira~Filho, and Storey}{Singer
  et~al\mbox{.}}{2014}]%
        {singer2014software}
\bibfield{author}{\bibinfo{person}{Leif Singer}, \bibinfo{person}{Fernando
  Figueira~Filho}, {and} \bibinfo{person}{Margaret-Anne Storey}.}
  \bibinfo{year}{2014}\natexlab{}.
\newblock \showarticletitle{Software engineering at the speed of light: how
  developers stay current using twitter}. In
  \bibinfo{booktitle}{\emph{Proceedings of the 36th International Conference on
  Software Engineering}}. ACM, \bibinfo{pages}{211--221}.
\newblock


\bibitem[\protect\citeauthoryear{Storey, Treude, van Deursen, and Cheng}{Storey
  et~al\mbox{.}}{2010}]%
        {storey2010impact}
\bibfield{author}{\bibinfo{person}{Margaret-Anne Storey},
  \bibinfo{person}{Christoph Treude}, \bibinfo{person}{Arie van Deursen}, {and}
  \bibinfo{person}{Li-Te Cheng}.} \bibinfo{year}{2010}\natexlab{}.
\newblock \showarticletitle{The impact of social media on software engineering
  practices and tools}. In \bibinfo{booktitle}{\emph{Proceedings of the FSE/SDP
  workshop on Future of software engineering research}}. ACM,
  \bibinfo{pages}{359--364}.
\newblock


\bibitem[\protect\citeauthoryear{Sun, Huang, and Kroening}{Sun
  et~al\mbox{.}}{2018a}]%
        {sun2018testing}
\bibfield{author}{\bibinfo{person}{Youcheng Sun}, \bibinfo{person}{Xiaowei
  Huang}, {and} \bibinfo{person}{Daniel Kroening}.}
  \bibinfo{year}{2018}\natexlab{a}.
\newblock \showarticletitle{Testing deep neural networks}.
\newblock \bibinfo{journal}{\emph{arXiv preprint arXiv:1803.04792}}
  (\bibinfo{year}{2018}).
\newblock


\bibitem[\protect\citeauthoryear{Sun, Wu, Ruan, Huang, Kwiatkowska, and
  Kroening}{Sun et~al\mbox{.}}{2018b}]%
        {sun2018concolic}
\bibfield{author}{\bibinfo{person}{Youcheng Sun}, \bibinfo{person}{Min Wu},
  \bibinfo{person}{Wenjie Ruan}, \bibinfo{person}{Xiaowei Huang},
  \bibinfo{person}{Marta Kwiatkowska}, {and} \bibinfo{person}{Daniel
  Kroening}.} \bibinfo{year}{2018}\natexlab{b}.
\newblock \showarticletitle{Concolic testing for deep neural networks}. In
  \bibinfo{booktitle}{\emph{Proceedings of the 33rd ACM/IEEE International
  Conference on Automated Software Engineering}}. ACM,
  \bibinfo{pages}{109--119}.
\newblock


\bibitem[\protect\citeauthoryear{Tian, Pei, Jana, and Ray}{Tian
  et~al\mbox{.}}{2018a}]%
        {tian2018deeptest}
\bibfield{author}{\bibinfo{person}{Yuchi Tian}, \bibinfo{person}{Kexin Pei},
  \bibinfo{person}{Suman Jana}, {and} \bibinfo{person}{Baishakhi Ray}.}
  \bibinfo{year}{2018}\natexlab{a}.
\newblock \showarticletitle{Deeptest: Automated testing of
  deep-neural-network-driven autonomous cars}. In
  \bibinfo{booktitle}{\emph{Proceedings of the 40th international conference on
  software engineering}}. ACM, \bibinfo{pages}{303--314}.
\newblock


\bibitem[\protect\citeauthoryear{Tian, Pei, Jana, and Ray}{Tian
  et~al\mbox{.}}{2018b}]%
        {tiandeeptest2018}
\bibfield{author}{\bibinfo{person}{Yuchi Tian}, \bibinfo{person}{Kexin Pei},
  \bibinfo{person}{Suman Jana}, {and} \bibinfo{person}{Baishakhi Ray}.}
  \bibinfo{year}{2018}\natexlab{b}.
\newblock \showarticletitle{Deeptest: {Automated} testing of
  deep-neural-network-driven autonomous cars}. In
  \bibinfo{booktitle}{\emph{Proceedings of the 40th international conference on
  software engineering}}. \bibinfo{publisher}{ACM}, \bibinfo{pages}{303--314}.
\newblock


\bibitem[\protect\citeauthoryear{Wan, Xia, Lo, and Murphy}{Wan
  et~al\mbox{.}}{2019}]%
        {wan2019does}
\bibfield{author}{\bibinfo{person}{Zhiyuan Wan}, \bibinfo{person}{Xin Xia},
  \bibinfo{person}{David Lo}, {and} \bibinfo{person}{Gail~C Murphy}.}
  \bibinfo{year}{2019}\natexlab{}.
\newblock \showarticletitle{How does Machine Learning Change Software
  Development Practices?}
\newblock \bibinfo{journal}{\emph{IEEE Transactions on Software Engineering}}
  (\bibinfo{year}{2019}).
\newblock


\bibitem[\protect\citeauthoryear{Wong and Bressler}{Wong and Bressler}{2016}]%
        {wong2016artificial}
\bibfield{author}{\bibinfo{person}{Tien~Yin Wong} {and} \bibinfo{person}{Neil~M
  Bressler}.} \bibinfo{year}{2016}\natexlab{}.
\newblock \showarticletitle{Artificial intelligence with deep learning
  technology looks into diabetic retinopathy screening}.
\newblock \bibinfo{journal}{\emph{Jama}} \bibinfo{volume}{316},
  \bibinfo{number}{22} (\bibinfo{year}{2016}), \bibinfo{pages}{2366--2367}.
\newblock


\bibitem[\protect\citeauthoryear{Xie, Ma, Juefei-Xu, Xue, Chen, Liu, Zhao, Li,
  Yin, and See}{Xie et~al\mbox{.}}{2019}]%
        {xie2019deephunter}
\bibfield{author}{\bibinfo{person}{Xiaofei Xie}, \bibinfo{person}{Lei Ma},
  \bibinfo{person}{Felix Juefei-Xu}, \bibinfo{person}{Minhui Xue},
  \bibinfo{person}{Hongxu Chen}, \bibinfo{person}{Yang Liu},
  \bibinfo{person}{Jianjun Zhao}, \bibinfo{person}{Bo Li},
  \bibinfo{person}{Jianxiong Yin}, {and} \bibinfo{person}{Simon See}.}
  \bibinfo{year}{2019}\natexlab{}.
\newblock \showarticletitle{DeepHunter: a coverage-guided fuzz testing
  framework for deep neural networks}. In \bibinfo{booktitle}{\emph{Proceedings
  of the 28th ACM SIGSOFT International Symposium on Software Testing and
  Analysis}}. ACM, \bibinfo{pages}{146--157}.
\newblock


\bibitem[\protect\citeauthoryear{Zhang, Harman, Ma, and Liu}{Zhang
  et~al\mbox{.}}{2019b}]%
        {zhang2019machine}
\bibfield{author}{\bibinfo{person}{Jie~M Zhang}, \bibinfo{person}{Mark Harman},
  \bibinfo{person}{Lei Ma}, {and} \bibinfo{person}{Yang Liu}.}
  \bibinfo{year}{2019}\natexlab{b}.
\newblock \showarticletitle{Machine Learning Testing: Survey, Landscapes and
  Horizons}.
\newblock \bibinfo{journal}{\emph{arXiv preprint arXiv:1906.10742}}
  (\bibinfo{year}{2019}).
\newblock


\bibitem[\protect\citeauthoryear{Zhang, Gao, Ma, Lyu, and Kim}{Zhang
  et~al\mbox{.}}{2019a}]%
        {zhangempirical}
\bibfield{author}{\bibinfo{person}{Tianyi Zhang}, \bibinfo{person}{Cuiyun Gao},
  \bibinfo{person}{Lei Ma}, \bibinfo{person}{Michael~R Lyu}, {and}
  \bibinfo{person}{Miryung Kim}.} \bibinfo{year}{2019}\natexlab{a}.
\newblock \showarticletitle{An Empirical Study of Common Challenges in
  Developing Deep Learning Applications}.
\newblock  (\bibinfo{year}{2019}).
\newblock


\bibitem[\protect\citeauthoryear{Zhang, Chen, Cheung, Xiong, and Zhang}{Zhang
  et~al\mbox{.}}{2018a}]%
        {zhangempirical2018}
\bibfield{author}{\bibinfo{person}{Yuhao Zhang}, \bibinfo{person}{Yifan Chen},
  \bibinfo{person}{Shing-Chi Cheung}, \bibinfo{person}{Yingfei Xiong}, {and}
  \bibinfo{person}{Lu Zhang}.} \bibinfo{year}{2018}\natexlab{a}.
\newblock \showarticletitle{An empirical study on {TensorFlow} program bugs}.
  In \bibinfo{booktitle}{\emph{Proceedings of the 27th {ACM} {SIGSOFT}
  {International} {Symposium} on {Software} {Testing} and {Analysis}}}.
  \bibinfo{publisher}{ACM}, \bibinfo{pages}{129--140}.
\newblock


\bibitem[\protect\citeauthoryear{Zhang, Liu, and Wang}{Zhang
  et~al\mbox{.}}{2018b}]%
        {zhang2018road}
\bibfield{author}{\bibinfo{person}{Zhengxin Zhang}, \bibinfo{person}{Qingjie
  Liu}, {and} \bibinfo{person}{Yunhong Wang}.}
  \bibinfo{year}{2018}\natexlab{b}.
\newblock \showarticletitle{Road extraction by deep residual u-net}.
\newblock \bibinfo{journal}{\emph{IEEE Geoscience and Remote Sensing Letters}}
  \bibinfo{volume}{15}, \bibinfo{number}{5} (\bibinfo{year}{2018}),
  \bibinfo{pages}{749--753}.
\newblock


\end{thebibliography}

\end{document}